\documentstyle[emulateapj,psfig]{article}
\newcommand{\beq}{\begin{equation}}
\newcommand{\eeq}{\end{equation}}
\newcommand{\bdm}{\begin{displaymath}}
\newcommand{\edm}{\end{displaymath}}
\newcommand{\bfig}{\begin{figure}}
\newcommand{\efig}{\end{figure}}
\newcommand{\msun}{\rm M_{\odot}}
\def\kms{\,{\rm km\,s^{-1}}}

\def\pcc{\,{\rm cm}^{-3}}

\def\etal{{et al.~}}
\def\ie{{\frenchspacing i.e. }}

\def\gtsima{$\; \buildrel > \over \sim \;$}
\def\ltsima{$\; \buildrel < \over \sim \;$}
\def\prosima{$\; \buildrel \propto \over \sim \;$}
\def\gsim{\lower.5ex\hbox{\gtsima}}
\def\lsim{\lower.5ex\hbox{\ltsima}}
\def\simgt{\lower.5ex\hbox{\gtsima}}
\def\simlt{\lower.5ex\hbox{\ltsima}}
\def\simpr{\lower.5ex\hbox{\prosima}}
\def\v{\mbox{\boldmath{$v$}}}
\def\g{\mbox{\boldmath{$g$}}}
\def\lta{\lsim}
\def\gta{\gsim}
\def\vir{{\rm vir}}
\def\lya{Ly$\alpha\ $}

\lefthead{Mori, Ferrara, \& Madau}
\righthead{Pregalactic Outflows}
\submitted{submitted to the ApJ}
\begin{document}

\title{EARLY METAL ENRICHMENT BY PREGALACTIC OUTFLOWS: II. THREE--DIMENSIONAL
SIMULATIONS OF BLOW--AWAY}

\author{Masao Mori\altaffilmark{1,2}, Andrea Ferrara\altaffilmark{2,3},
and Piero Madau\altaffilmark{3,4}}
\altaffiltext{1}{Institute of Natural Science, Senshu University, 
Kawasaki-shi, Kanagawa-ken, 214-8580, Japan.}
\altaffiltext{2}{Center for Computational Physics, University of Tsukuba, 
Tsukuba-shi, Ibaraki-ken, 305-8577, Japan.}
\altaffiltext{3}{Osservatorio Astrofisico di Arcetri, Largo Enrico Fermi 5, 
50125 Firenze, Italy.}
\altaffiltext{4}{Department of Astronomy and Astrophysics, University of
California, Santa Cruz, CA 95064.}

\begin{abstract}
Supernova (SN)--driven pregalactic outflows may be an efficient mechanism
for distributing the product of stellar nucleosynthesis over large cosmological
volumes prior to the reionization epoch. Here we present results from 
three--dimensional numerical simulations of the dynamics of SN--driven 
bubbles as they propagate through and escape the grasp of subgalactic halos 
with masses $M=10^8\,h^{-1}\,\msun$ at redshift $z=9$. Halos in this mass 
range are characterized by very short dynamical timescales (and even shorter 
gas cooling times) and may therefore form stars in a rapid but 
intense burst before SN `feedback' quenches further star formation.
The hydrodynamic simulations use a nested grid method to follow the evolution 
of explosive multi--SN events operating on the characteristic timescale of a 
few $\times 10^7\,$yr, the lifetime of massive stars. The results confirm 
that, if the star formation efficiency of subgalactic halos is $\lta 10\%$, 
a significant fraction of the halo gas will be lifted out of the potential 
well (`blow--away'), shock the intergalactic medium, and pollute it with 
metal--enriched material, a scenario recently advocated 
by Madau, Ferrara, \& Rees (2001). The volume filling factor of the ejecta is
of order unity. Depending on the stellar distribution,
we find that less than 30\% of the available SN energy gets converted 
into kinetic energy of the blown away material, the remainder 
being radiated away. It appears that mechanical feedback is 
less efficient than expected from simple energetic arguments, as 
off--nuclear SN explosions drive inward--propagating shocks that tend to 
collect and pile up cold gas in the central regions of the host halo. 
Low--mass galaxies at early epochs may survive multiple SN events and 
continue forming stars.

\end{abstract}

\section{Introduction}      
In currently popular hierarchical clustering scenarios for the formation of cosmic
structures -- all variants of the cold dark matter (CDM) cosmogony -- the assembly
of galaxies is a bottom--up process in which large systems result from the merging 
of smaller subunits. In these theories dark matter halos with masses $M=10^8\,
h^{-1}\,\msun$ collapse at $z\approx 10$ from 2--$\sigma$ density fluctuations: 
the gas 
infalling along with the dark matter perturbation is shock--heated to the virial 
temperature $T_\vir=2\times 10^4\,$K, condenses rapidly due to 
atomic line cooling, and becomes self--gravitating. Massive stars then form 
with some initial mass function (IMF), synthesize heavy--elements, and 
explode as supernovae (SNe) after a few $\times 10^7\,$yr, enriching the 
surrounding medium.
These subgalactic stellar systems, possibly aided by a population of accreting 
black holes in their nuclei and/or by an earlier generation of stars in even 
smaller halos (`minihalos' with virial temperatures of only a few hundred kelvins, 
where collisional excitation of molecular hydrogen is the main coolant), 
are believed to have generated 
the ultraviolet radiation and mechanical energy that reheated and 
reionized the universe (see Loeb \& Barkana 2001 for a recent review).

It is a simple expectation of the above scenario that the energy deposition 
by SNe in the shallow potential wells of subgalactic systems may have two 
main effects, depending on the efficiency with which halo gas can cool and 
fragment into clouds and then into massive stars: (i) the disruption of the 
newly formed object, \ie the most violent version of the so--called `stellar
feedback' (Larson 1974; Dekel \& Silk 1986; Mori \etal 1997; 
Mori, Yoshii, \& Nomoto 1999; MacLow \& Ferrara 1999; 
D'Ercole \& Brighenti 1999; Murakami \& Babul 1999; Ferrara \& Tolstoy 2000; Ciardi \etal 2000); 
and (ii) the blow--away of metal--enriched baryons from 
the host (dwarf) galaxy, causing the pollution of the IGM at early times
(Tegmark, Silk, \& Evrard 1993; Voit 1996; Nath \& Trentham 1997; Gnedin \& 
Ostriker 1997; Ferrara, Pettini, \& Shchekinov 2000; Madau, Ferrara, \& Rees
2001, hereafter Paper I). The well--established
existence of heavy elements like carbon, nitrogen, and silicon in the \lya
forest clouds at $z=3-3.5$ may be the best evidence for such an
early episode of pregalactic star formation and outflows (Songaila 1997; 
Ellison \etal 2000). Stellar feedback and galactic outflows at high--redshifts 
may also temporarily halt or delay the formation of dwarf galaxies (Scannapieco
\& Broadhurst 2001; Scannapieco, Ferrara, \& Broadhurst 2000), affect the 
thermal state of the IGM (Madau 2000;
Theuns, Mo, \& Schaye 2001; Cen \& Bryan 2001), and play a crucial role during the 
reionization epoch (Ciardi \etal 2000; Bruscoli \etal 2000).

This is the second paper in a series aimed at a detailed study of stellar feedback
and SN--driven pregalactic outflows at early epochs. In Paper I 
it was shown that the observed narrow Doppler widths and 
inferred large filling factor of chemically enriched, low--density \lya forest 
clouds may point to a more uniform (i.e. `early') rather than in--situ (i.e. 
`late') cosmic metal pollution. 
Whilst outflows of metal--rich gas are directly observed in 
Lyman--break galaxies at $z\approx 3$ (Pettini \etal 2000), most of this gas may 
not leave the immediate surroundings of these deep gravitational
potential wells. We argued that, if intergalactic metals were actually
dispersed {\it over large cosmological volumes} from massive galaxy halos at late 
times, such a delayed epoch of galactic
``super--winds'' would have severely perturbed the IGM (since the kinetic energy of
the ejecta is absorbed by intergalactic gas), raising it to a much higher adiabat 
than expected from photoionization, and producing large spatial variations of the 
baryons relative to the dark matter. This would make the success of cosmological
hydrodynamical simulations in matching the overall observed properties of \lya 
absorption systems largely coincidental. Alternatively, the observations 
could be best explained if the
ejection of heavy elements at velocities exceeding the small escape speed of
numerous subgalactic systems were to take place at very high redshifts. This is 
because hot enriched
material cools more efficiently at these early epochs, and the expansion of
SN--driven bubbles into a dense IGM (pre--photoionized by the same massive 
stars that later explode as SNe) would be halted by the external pressure. In this
scenario, any residual peculiar velocity would have been redshifted away by $z=3$, 
the \lya forest would be hydrodynamically `cold', and the intergalactic baryons 
would have relaxed again under the influence of dark matter gravity.

In Paper I the most efficient pollutant of the IGM on large scales were 
identified as subgalactic systems with masses $\sim 10^8\,h^{-1}\,\msun$ at $z\sim
10$, when large number of them grow non--linear and collapse.
We showed how stellar feedback in these pregalactic halos (with virial temperatures 
$T_\vir=2\times 10^4\,$K, i.e. at the peak of the cooling curve for primordial gas) 
may be an inherently different process than that often invoked 
to prevent the `overcooling' of baryons in the 
central regions of larger systems (White \& Frenk 1991; Cole \etal 1994; Navarro 
\& Steinmetz 1997, 2000). This is because such fragments are characterized by 
very short dynamical timescales (and even shorter gas cooling times): throughout 
the halo the accreted baryons condense immediately due to atomic hydrogen cooling, 
and the supply of cold gas for star formation is only limited by the infall rate. 
They are therefore expected to go through a rapid but intense star forming phase, 
since one cannot
assume {\it instantaneous feedback} from SNe in systems with such a short dynamical 
timescale. Lacking the ability  to store efficiently the corresponding energy input 
in turbulent  and magnetic forms or by relaxing to a multiphase medium, 
these system have very limited resources to sustain a self--regulated star
formation cycle (cf. Efstathiou 2000).

In Paper I we used the thin shell approximation to model the evolution of 
SN--driven
bubbles as they propagate through and blow out from subgalactic halos.
We assumed spatial coherency among SN events and temporal coherency among their
progenitors, i.e. all 
explosions were assumed to take place at the center of the halo, and the 
progenitor stellar population to form coevally, on a timescale short compared to 
the lifetime of massive stars. In this paper we present results from 
three--dimensional
hydrodynamic simulations of the dynamics of SN--driven bubbles in 
subgalactic halos using a hierarchical nested grid technique.
Besides accounting for the mass--dependent main sequence evolutionary timescale
of massive stars (as in Paper I), which causes a temporal spread of SN 
explosions, we also explore the 
role of different spatial distributions of the explosion sites inside the 
galaxy, thus relaxing the assumption that all SNe occur at the center.
The crucial questions we try to give a quantitative answer in this paper are: 
How efficient mechanical feedback really is? Can small halos 
survive multiple SN events and continue forming stars or are they blown away 
and end their life as `naked stellar clusters'? How far from the production 
sites can metals be ejected into the IGM?  

\section{Model Assumptions}

We will model the structural properties of subgalactic systems following 
Paper I. We shall assume that 
virialized dark matter halos, formed through hierarchical
clustering, have a universal (spherically averaged) Navarro--Frenk--White (1997,
hereafter NFW) density profile,
\beq
\label{rhoh}
\rho_{\rm DM}(r)={\rho_{\rm crit}\,\delta_c\over cx (1+cx)^2}, 
\eeq
where $x\equiv r/r_\vir$, $r_\vir$ is the virial radius of the system, \ie 
the radius of the sphere encompassing a mean overdensity of 200, $c$ is the
halo concentration parameter, $\delta_c=(200/3)c^3/F(c)$ is a characteristic
overdensity, $\rho_{\rm crit}=3 H^2/(8\pi G)$ the critical density of the
universe\footnote{Unless otherwise stated, we will assume throughout this paper an 
Einstein--de Sitter cosmology with $\Omega_M=1$, $\Omega_\Lambda=0$, 
and a Hubble constant of $H_0=100\,h$~km~s$^{-1}$~Mpc$^{-1}$. The 
numerical simulations are run with $h=0.5$.}\, at redshift 
$z$, and
\beq
F(t)\equiv \ln(1+t)-{t\over 1+t}.
\eeq
The total within the virial radius is $M=(4\pi/3)200\rho_{\rm
crit}r_\vir^3$, and the `typical' concentration parameter of 
halos at $z=10$ with $M=10^8\,h^{-1}\,\msun$ is $c=5$ (Paper I). 
Note that high--resolution N--body simulations by Bullock \etal (2001) indicates 
that high--redshift halos are actually less concentrated that expected from the 
NFW 
prediction. In this case we may be overestimating the escape speed from
subgalactic systems.

For our prototypical pollutor (below the mass scale given above 
the main coolant is molecular hydrogen and halos are strongly affected by
radiative feedback, while above this mass scale cooling is less rapid, halos
are more rare, and metals are more efficiently retained by the deeper potential
well, see detailed discussion in Paper I), the characteristic virial radius,    
virial temperature, circular velocity at the virial radius, and escape 
velocity at the center are
\beq
r_\vir=0.75\,{\rm kpc}\,M_8^{1/3}\, h^{-1}\,\left({1+z\over 10}\right)^{-1},
\eeq
\beq
T_\vir=2\times 10^4\,{\rm K}\,M_8^{2/3}\,\left({1+z\over 10}\right),
\eeq
\beq
V_c=24\,{\kms}\,M_8^{1/3}\,\left({1+z\over 10}\right)^{1/2},
\eeq
\beq
v_e(0)=77 \,{\kms}\,M_8^{1/3}\,\left({1+z\over 10}\right)^{1/2},
\eeq
respectively, where $M_8$ is the halo mass in units of $10^8\,h^{-1}\,\msun$. 
Here we have set the mean molecular weight $\mu$ to 0.59, appropriate 
for a fully ionized primordial gas. At the virial radius 
$v_e(r_\vir)=0.62\,v_e(0)$. If baryons virialize in the dark matter halo 
to an isothermal distribution, they will be shock--heated to the virial temperature
and settle down to a density profile
$$
\ln\rho(r)=\ln\rho_0\,-{\mu m_p\over 2kT_\vir}[v_e^2(0)-
v_e^2(r)] \nonumber\\
$$
\beq
~~~~~~~~~~~=\ln\rho_0\,-{2c\over F(c)}\left[1-{\ln(1+cx)\over 
cx}\right],
\label{rhog}
\eeq
where $m_p$ is the proton mass. The central gas density 
$\rho_0$ is determined by the condition that the total baryonic mass fraction 
within the virial radius is equal to $\Omega_b$ initially. Adopting 
$\Omega_b h^2=0.019$ (Burles \& Tytler 1998), one gets $\rho_0=840\,h^{-2}\,
\rho_{\rm crit}=1.6\times 10^{-23}\,$ gr cm$^{-3}$ at $z=9$. The halo gas density
at $r_\vir$ is $\rho(r_\vir)=0.00144\rho_0=2.3\times 10^{-26}$ gr cm$^{-3}$. 
As this  
is about 60 times higher at $(1+z)=10$ than the IGM density, to avoid unphysical 
effects due to this jump we have allowed the two distributions to merge trough 
a hyperbolic tangential transition of width $0.05\,r_\vir$.

Numerical N--body/hydrodynamics simulations of structure formation in CDM 
cosmologies 
have provided in the last few years a definite picture for the topology of the IGM,
one of an interconnected network of sheets and filaments with virialized systems 
located at their points of intersection. Intergalactic gas will be infalling onto 
a galaxy along the filaments of this ``cosmic web'', at approximately the halo
escape velocity. Our simulations do not include the effect of this non--spherical 
infall on SN--driven outflows. Note that, for the halos/outflows under consideration, 
the escape speed at the virial radius is $v_e(r_{\rm vir})=48\,
\kms$, while the outflow velocity is close to $150\,\kms$ for a star formation 
efficiency of 10\% (see Fig. 7 of Paper I). In this case the pressure due to 
infalling material is 9 times smaller than the ram pressure of the 
expanding shell, and can be neglected in first approximation. As the blast 
propagates into the IGM, however, it may preferentially enrich the voids in between 
the denser filaments. SN--driven outflows in a realistic cosmological density field
will be the subject of a subsequent paper in this series.

\subsection{Supernova progenitors} 

As shown in Paper I, for halo masses in the range $10^8\,h^{-1}\lta M\lta 10^{10}\,
h^{-1}\,\msun$ at $(1+z)=10$, the cooling time is shorter than the dynamical
timescale everywhere in the halo: infalling gas never comes to hydrostatic 
equilibrium, but collapses to the center at the free--fall rate. 
In our prototypical halo, the free--fall time is $10^7$~yr at $r=0.1\,r_\vir$, 
increasing to $8\times 10^7$~yr at $r_\vir$ (Paper I). The cooling time for
gas at $T_\vir$ is always shorter than the free--fall time by more than two
orders of magnitude. Mechanical energy will be injected by SNe only after a 
few times $10^7\,$ yr: at this stage SN--driven bubbles will propagate into the 
halo
quenching further star formation, and the conversion of cold gas into stars
will be limited by the increasing fractional volume occupied by supernova
remnants. 

Before SN feedback starts operating, some fraction $f_\star$  of the halo 
gas may be able to cool, fragment, and form stars. In general, cooling
is efficient when the cooling time $t_{\rm cool}=1.5nkT/(n_{\rm H}^2\Lambda)$ is much 
shorter than the local gas--dynamical timescale $t_{\rm dyn}=(4\pi G\rho)^{-1/2}$, 
where $\Lambda$ is the radiative cooling funtion.
This condition implies that the energy deposited by gravitational 
contraction cannot balance radiative losses; as a consequence, the temperature 
decreases with increasing density, the cloud cools and then fragments. At any given 
time, fragments form on a scale that is small enough to ensure pressure equilibrium 
at the corresponding temperature, \ie the Jeans length scale.

Small pregalactic systems at early times, like the ones under consideration, 
may then be expected 
to consume a fair fraction of their cold gas in a single burst of star
formation. Note that the highest 
efficiency estimated in nearby star--forming regions is about 30\% for the 
Ophiuchi dark cloud (Wilking \& Lada 1983; Lada \& Wilking 1984), and that 
the UV radiation from massive stars may further inhibit cooling and delay 
the collapse of mildly overdense regions (Nishi \& Tashiro 2000). 
In the following we will adopt an illustrative value of $f_\star=10$\% for 
the star formation efficiency. This is consistent with the stellar--to--baryonic 
mass ratio observed today, and implies that only a few percent of the 
present--day stellar mass would form at these 
early epochs (see discussion in \S\,6). The ensuing SNe (assuming a Salpeter
initial mass function) would pollute the IGM to a mean metallicity of order 0.3\% 
solar, comparable to the levels observed in the \lya forest at redshift 3.
Given our poor understanding of star formation, however, $f_\star$ should really 
be considered at this stage as a free parameter of the theory.

The halo under study has a baryonic mass $M_b=\Omega_b M=0.019\,h^{-2}\times  
10^8 h^{-1}\,\msun=1.9\times 10^6 h^{-3}\,\msun$. Its total stellar mass is then
$M_\star= f_\star M_b= 1.5\times 10^6\,\msun$ ($h=0.5$). 
An amount $M_\star$ of baryons is subtracted from the total gas mass;
the rest of the gas is assumed to keep the same initial distribution. 
We assume that the initial stellar mass function (IMF) can be approximated by 
a Salpeter power--law with lower and upper
mass cutoffs equal to $m_l=0.1\,\msun$ and $m_u=120\,\msun$, respectively.
Note that considerable uncertainties still remain on the characteristic mass of
the first luminous objects. Numerical simulations of the fragmentation of 
primordial clouds with masses $\gta$ a few $\times 10^5\,\msun$ in 
hierarchical cosmogonies have suggested that the IMF of the very first,
zero--metallicity stars forming at $z\gta 20$ may be extremely top--heavy 
(Bromm, Coppi, \& Larson 1999, 2001a; Abel, Bryan, \& Norman 2000), perhaps
giving origins to a population of pregalactic massive black holes 
(Madau \& Rees 2001).  
As this feature appears linked to primordial H$_2$ chemistry and cooling, at 
the later epochs of interest here and in more massive halos where gas condense 
due to atomic line cooling, it is plausible that a 
`second' generation of stars may be able to form with an IMF that is less 
biased towards very high stellar masses (Bromm \etal 2001b).

In our simulation, the mass of each star is determined by randomly sampling the 
IMF until the total stellar mass is equal to $M_\star$; 
SN progenitors are then identified as those stars more massive than 8~M$_\odot$. 
We find that a total number of $N_t=11170$ SNe are produced in the system,
out of $4.24\times 10^6$ stars (average stellar mass = 0.35 M$_\odot$)
yielding a SN every 134 M$_\odot$ of stars formed. Each massive star 
(with mass $m$ in units of $\msun$) explodes after a main sequence lifetime of 
\begin{eqnarray}
\log t_{\rm ms}{\rm ~(yr)} = 10.025-3.559\log m +0.898 (\log m)^2.
\end{eqnarray}
This is a fit to a compilation of the available data in the literature 
(Schaller \etal 1992; Vacca, Garmany \& Shull 1996; Schaerer \& de Koter 1997; 
Palla, private communication). In the local universe SNe are known not to 
occur at random location but rather to cluster into OB
associations (Heiles 1990). In nearby galaxies the luminosity 
function of OB associations is well approximated by a power--law
\begin{equation}
\label{dist}
{d{\cal N}_{\rm OB}\over dN} = A N^{-\beta},
\end{equation}
with $\beta \approx 2$ (McKee \& Williams 1997; Oey \& Clarke 1998).
Here ${\cal N}_{\rm OB}$ is the number of associations containing $N$ OB stars; 
the probability for a cluster of OB stars to host $N$ SNe is  then
$\propto N^{-2}$. We apply a grouping procedure according to the above 
distribution: this yields 90 OB associations, each containing a variable 
number of massive stars ranging from a few tens to up to 2000.
For numerical reasons to be discussed below, we set a lower  
limit to the number of stars in an association equal to 20.

As a final step we have to spatially locate the 90 associations 
in the halo. Our modelling of star formation is, unfortunately, limited by 
finite numerical resolution and the neglect of self--gravity.     
We have therefore decided to use a simple scheme 
that incorporates some gross observed features of star formation and
makes use of actual gas properties such as the local mass density.
We distribute SNe according to the gas density power--law introduced
by Schmidt (1959), \ie the number density of explosions is proportional 
to $\rho^\alpha$.
In starburst and spiral galaxies, the disk--averaged star formation rates and gas 
densities are well represented by a Schmidt law with $\alpha=1.4\pm 0.15$
(Kennicutt 1998). Numerous theoretical scenarios that produce a Schmidt law with 
$\alpha=1-2$ can be found in the literature (e.g. Silk 1997); since the 
cooling time is much shorter than the gas dynamical timescale in dense 
star--forming regions, the star formation rate is often taken to be 
proportional to $\rho/t_{\rm dyn}\propto \rho^{3/2}$. Similar phenomenological 
prescriptions for star formation are used in numerical simulations of
hierarchical galaxy formation (Katz 1992). In this paper we run two simulations,
corresponding to a relatively extended ($\alpha=1$) or more centrally concentrated 
($\alpha=2$) star formation volume (in the limit $\alpha\rightarrow \infty$ all 
SNe explode at the center). We will refer to them in the following as Case 1 
($\alpha=1$) and Case 2 ($\alpha=2$), respectively. 

It is important to notice that our assumption of an extended spherical distribution 
of SNe may be a poor one if the fragmentation of cooling gas freely--falling 
towards  the 
halo center is an inefficient process (as argued by Kashlinsky \& Rees 1983),
and star formation only occurs after cold halo gas actually settles down in a 
centrifugally--supported galactic disk. If an exponential disk with scale 
length $r_d$ forms, and the specific
angular momentum of the disk material is the same as that of the dark halo (treated
for simplicity as a singular isothermal sphere), then
angular momentum conservation fixes the collapse factor to $r_\vir/r_d=\sqrt{2}/
\lambda$, where $\lambda\approx 0.05$ is the spin parameter. Our fiducial halo/disk 
system at $z=9$ would then be characterized by a scale length of only 
$r_d=27\,h^{-1}\,$pc (Paper I). In this scenario, and with the resolution 
of our numerical 
simulations (see below), all SNe explosions would take place within a 
few central mesh cells. Also, a thin dense disk would hardly couple to the
outflow, and matter would be ejected perpendicularly to the disk (MacLow \& Ferrara 
1999). It is unclear, however, whether thin self--gravitating disks would actually 
form and/or survive in subgalactic fragments. If a disk of mass $M_d$
follows an isothermal vertical profile with a thermal speed $c_s=10\,\kms$,
typical of gas which is continuosly photoheated by stars embedded within the
disk itself, then its scale height at radius $r_d$ is $h/r_d=\sqrt{16e}\,
\lambda (M/M_d) (c_s/V_c)^2$, and the disk would be thick rather than thin.
Recent numerical simulations of the formation and fragmentation of primordial 
molecular clouds in CDM cosmologies, with realistic initial conditions,
do not show the presence of a disk, and form the first dense fragments close to 
the center of the halo (Abel \etal 2000). A rotationally--supported disk 
that fragments efficiently appears to only form as a consequence of 
idealized initial conditions, i.e. a top--hat isolated sphere that initially
rotate as a solid body (Bromm \etal 1999). 

\section{Preliminary Considerations}

Before discussing the detailed results of the numerical simulations,
it is important to provide some physical insights into the problem.   

\subsection{Effects of explosion location} 

In order for the gas to be ejected from the galaxy, a necessary       
condition implies that the initial energy provided by the supernovae has
to be larger than the gravitational one. The latter is given by   
\beq
\label{Eg}
E_g = {1\over 2} \int_V  d^3{\bf r}\rho(r) \phi(r) 
\eeq
where $\phi(r) = -GM(r)/r$ is the gravitational potential. With the gas density
distribution given by equation (\ref{rhog}), integration yields
$$
\label{Eg1}
E_g ={3\,V_c^2\,M\over 400\, F(c)}\int_0^1 dx\,x\,F(cx)\,\left({\rho\over 
\rho_{\rm crit}}\right)=0.01\,h^{-2}\,V_c^2 M
$$
\beq
~~~~~~~~=10^{52}\,h^{-3}\,{\rm erg}\left({1+z\over 10}\right)\,M_8^{5/3}.
\eeq
This energy must be compared with the total mechanical
energy injected by SN explosions, $E_{\rm SN}=N_t E_0=1.1 \times 10^{55}$~erg. 
The naive conclusion would be that 
this energy deposition, being roughly hundred times higher than the gas binding
energy, should produce a complete blow--away of the halo. There are two main
reasons, however, for why this is not actually the case: i) the conversion 
efficiency of $E_{\rm SN}$ into kinetic energy of the interstellar medium (ISM) 
is well below unity 
since radiative energy losses, particularly at the halo center where the baryon 
density is initially $n(0) \approx 10\pcc$, carry away a significant fraction 
of this energy; and ii) as the explosion sites are scattered within the host
galaxy, they can have very different effects. For example, explosions 
taking place in the outer halo will tend to expand outwards along the rapidly 
decreasing density gradient: the released energy will be eventually used to
energize the IGM. Off--nuclear SN explosions will 
drive inward--propagating shocks that pile up and compress the gas in 
the central regions, further increasing radiative losses.
Also, energy deposition by an association at 
the center may never accelerate material beyond the virial radius. 

\begin{figure}
\centerline{
\psfig{figure=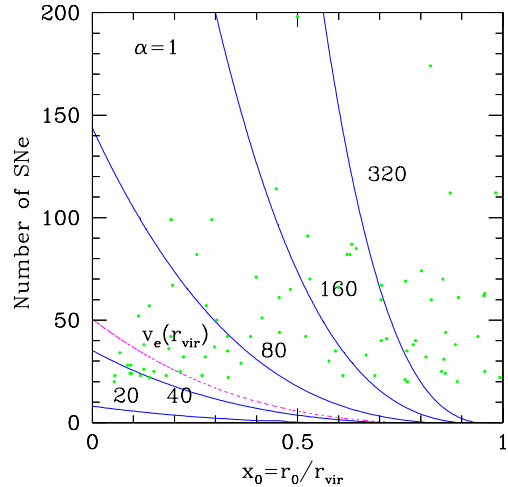,height=10cm}}
\caption{\footnotesize Isocontours in the $x_0--N$ plane of bubble shell 
velocity, $v_s(r_\vir)$ (in km~s$^{-1}$), at the virial radius. Here $N$ SNe are 
assumed to explode at a distance $x_0$ from the center (in units of the halo virial 
radius). Also shown ({\it dotted line}) is the isocontour corresponding to 
the escape speed $v_e(r_\vir)$. The points indicate the locus of the associations 
used in the numerical simulations for $\alpha=1$ (Case 1).}
\label{fig1}
\end{figure}

\begin{figure}
\centerline{
\psfig{figure=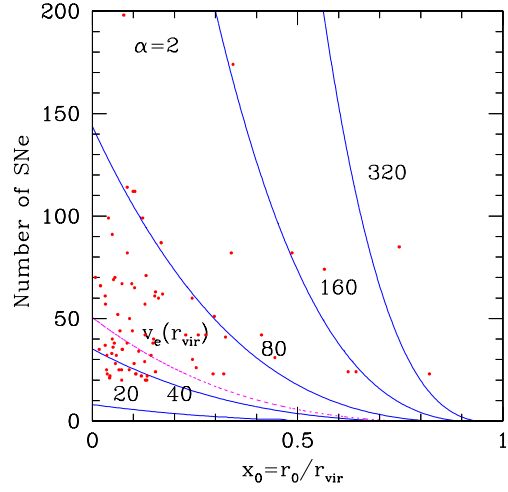,height=10cm}}
\caption{\footnotesize Same as Fig. \ref{fig1}, but for $\alpha=2$ (Case 2).}
\label{fig2}
\end{figure}

We can better quantify the last point by adopting the following zeroth order
approximation. We treat each association as a point explosion (in practice,
due to the spread of the various SN events, the ensuing superbubble is better
described by a wind solution, see Paper I). The evolution of a point explosion 
in a stratified medium can be treated by the thin--shell approximation. This 
solution accurately approximates the exact numerical results (see, e.g., 
MacLow \etal 1989) and can be obtained from dimensional analysis.
The shell velocity is $v_s(r)\simeq [P/\rho(r)]^{1/2}$, where the 
pressure $P$ is roughly equal to $E/X^3$: $E=N E_0$ is the energy injected 
by $N$ SNe in the association. The explosion is allowed to take place at 
different radial positions $x_0=r_0/r_\vir$; the bubble size along the radial 
coordinate is then $X=r_\vir (x-x_0)$. We can then calculate the value of 
$v_s(r_\vir)$, which can then be compared to the escape velocity $v_e(r_\vir)$ 
to get an idea  of the 
final fate (escape versus recapture) of the ejected gas. The results 
are shown in Figures \ref{fig1} and \ref{fig2} as  $v_s(r_\vir)$ 
isocontours in the plane identified by $x_0$ and $N$. Also
shown are the locations in that plane of the associations used in the 
numerical simulations for the two cases considered. 
It can be seen that, if the explosion occurs at the halo center ($x_0=0$),
a minimum of about 50 SNe is required in order to achieve shell velocities 
in excess of $v_e(r_\vir)$. This constraint is much weaker at (say) $x_0=0.7$, 
where even a single SN event can produce the same effect. Thus, every explosion 
occurring beyond this location will easily lead to mass loss from the galaxy. 
Note that in Case 1 a larger fraction (about 90\%) of SN events will give 
origin to significant mass loss, as they take place preferentially in the 
galactic outskirts. By contrast, in Case 2 about 1/3 of the OB associations 
generate only sub--escape speeds, i.e. $v_s(r_\vir)<v_e(r_\vir)$. 

\subsection{Energy-- and mass--carrying winds}

\begin{figure}
\centerline{
\psfig{figure=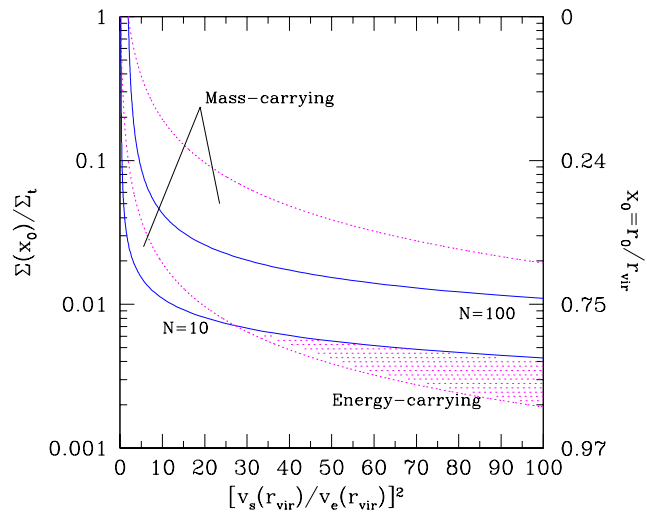,height=10cm}}
\caption{\footnotesize Halo gas column density $\Sigma(x_0)$ (normalized to 
the total gas column density through the halo $\Sigma_t$) vs. 
$[v_s(r_\vir)/v_e(r_\vir)]^2$ for two values of $N=10,100$ ({\it solid curves}).
The dotted lines depict the corresponding curves of constant kinetic energy 
flux, ${\cal E}_k(N)$. The shaded area indicates the parameter space
in which the wind is energy--carrying.    
}
\label{fig3}
\end{figure}

Explosions located in the halo outer regions produce higher 
velocities of the outflowing material. The swept--up mass, however, decreases
as the halo density is lower in the outskirts. In this case SNe basically vent 
energy (an energy--carrying wind) into the IGM, but little galactic mass. An 
accurate determination of the kinetic energy carried away, $E_k \propto M_s 
v_s^2$, can only be done numerically as it depends on the detailed mass loading 
of the propagating shells. To get some insight, we estimate the kinetic 
energy flux through $r_\vir$ as 
\beq
{\cal E}_k(x_0,N)=v_s^2(r_\vir) \Sigma(x_0) = {N E_0 r_\vir\over X^3 \rho} 
\int_{x_0}^1 dx \rho,  
\eeq
where $\Sigma(x_0)$ is the halo gas (mass) column density along the radial 
direction from $x_0$ to $r_\vir$. 
We then plot in Figure \ref{fig3} the column $\Sigma$ normalized to the total
column through the halo center, $\Sigma_t$, as a function of the square 
of the ratio $v_s(r_\vir)/v_e(r_\vir)$. 
If the kinetic energy flux were the same at every location one would find 
$\Sigma v_s^2(r_\vir)={\rm const}={\cal E}_k(N)$. As seen from Figure 
\ref{fig3}, however, the trend is rather different. 
For $N=10$ the curve drops very rapidly below the one for ${\cal E}_k(10)$:
this is because the increase in the shell velocity cannot compensate for 
the drop in mass as the explosion is displaced from the origin. It is
only in the outskirts ($x_0 > 0.75$) that the transition to a predominantly
energy--carrying wind occurs. In the $N=100$ case such inversion in not
seen, essentially because the mass decrease always dominate over the 
velocity increase.

In conclusion, if SNe occur in small $N\approx 20-30$ associations
(or are isolated),  either there is no mass--loss (if they are located 
at the center) or they mostly inject 
energy rather than mass into the IGM (if they are located in the outer halo). 
It is only when they are grouped together in larger associations that 
they will be able to eject both mass and energy into the intergalactic space.

\subsection{Porosity and amplification effect}

Based on the results shown in Figures \ref{fig1} and \ref{fig2}, Case 2 should 
lead to a smaller mass loss from our subgalactic halo. This, however, may  
not be necessarily the case, as ``amplification'' effects can be important 
in the central regions. Due the high stellar density at $x<0.2$ (say),
overlapping of the hot bubble interiors does occur if cooling is not too 
strong. In this case, the energy deposited at the various 
explosion sites can add and act coherently on the same shell, leading to an 
amplification of the pressure force and to higher shell velocities. In other 
words, the porosity of the hot gas, $Q$, (its volume filling factor being 
equal to $1-e^{-Q}$) is a key parameter, as already pointed out
by Silk (1997) and Efstahtiou (2000) in the context of feedback-regulated
star formation in galaxy disks. The effect of the interaction and merging of 
individual bubbles will be discussed in the framework of the 
numerical simulations discussed below.

\section{Numerical Method}  

The evolution of the gas is described by the three--dimensional 
hydrodynamic equations for a perfect fluid in Cartesian geometry. 
The continuity equation (including a term due to mass ejection from SNe, 
$\dot{\rho_{\rm s}}$), the momentum equation with the
gravitational acceleration $\g=(g_x, g_y, g_z)^{\rm T}$ , and 
the thermal energy equation associated with the rates of cooling 
$\Lambda$ and SNe heating $\Gamma_{\rm s}$, can be written in compact form as 
\begin{eqnarray}
\frac{\partial \bf U}{\partial t}
	+\frac{\partial \bf F}{\partial x}
	+\frac{\partial \bf G}{\partial y}
	+\frac{\partial \bf H}{\partial z}
	=\bf S
\label{compact}
\end{eqnarray}
with
\begin{eqnarray}
{\bf U}
 &=&(\rho    , \rho v_x    , \rho v_y    , \rho v_z    , 
 \rho     e)^{\rm T},\\
{\bf F}
 &=&(\rho v_x, \rho v^2_x+p, \rho v_x v_y, \rho v_x v_z, 
 \rho v_x h_e)^{\rm T},\\
{\bf G}
 &=&(\rho v_y, \rho v_x v_y, \rho v^2_y+p, \rho v_y v_z, 
 \rho v_y h_e)^{\rm T},\\
{\bf H}
 &=&(\rho v_z, \rho v_x v_z, \rho v_y v_z, \rho v^2_z+p, 
 \rho v_z h_e)^{\rm T},\\
{\bf S}
 &=&(\dot{\rho}_{\rm s}, \rho g_x, \rho g_y, \rho g_z, 
	\rho \v\cdot\g+\Gamma_{\rm s}-\Lambda)^{\rm T}.
\end{eqnarray}
Here ${\bf U}$ is a state vector of conserved quantities,
${\bf F, G,}$ and ${\bf H}$ are the corresponding flux vectors, and
${\bf S}$ is the source--term vector that includes sources and sinks
of conserved quantities, such as heating and cooling terms and 
the gravitational acceleration. Also, $\rho$ is the gas density, 
$\v=(v_x, v_y, v_z)^{\rm T}$ is the gas velocity vector, 
\begin{eqnarray}
e = \varepsilon+\frac{1}{2} |\v|^2 = h_e-\frac{p}{\rho}
\end{eqnarray}
is the specific total energy,  
$h_e$ is the enthalpy, and the internal energy $\varepsilon$ is related to 
the gas pressure by
\begin{eqnarray}
p = ( \gamma -1 ) \rho \, \varepsilon,
\end{eqnarray}
where $\gamma$ (=5/3) is the adiabatic index.

The equations are solved by a finite volume scheme with 
operator splitting, which is based on the AUSMDV described by 
Wada \& Liou (1997). Liou \& Steffen (1993) developed a 
remarkably simple upwind flux vector splitting scheme called 
`advection upstream splitting method' (AUSM). It treats the 
convective and pressure terms of the flux function separately.
The AUSMDV has a blending form of AUSM and flux difference, 
and improves the robustness of AUSM in dealing with the 
collision of strong shocks. We extended it to third--order 
spatial accuracy using MUSCL (van Leer 1977) with a total 
variation diminishing limiter proposed by Arora \& Roe (1997).
Since this scheme has a great advantage due to the reduction 
of numerical viscosity, fluid interfaces are sharply preserved 
and small--scale features can be resolved as in the `piecewise 
parabolic method' (PPM) of Woodward \& Colella (1984). The AUSMDV 
scheme is, however, simpler and has a lower computational costs than PPM. The 
code has passed successfully several tests and handles very well
weak and high Mach number shocks. In the Appendix we present 
the results for two such tests, a standard one--dimensional shock tube and a 
point explosion in three dimensions.

\begin{figure*}
\centerline{
\psfig{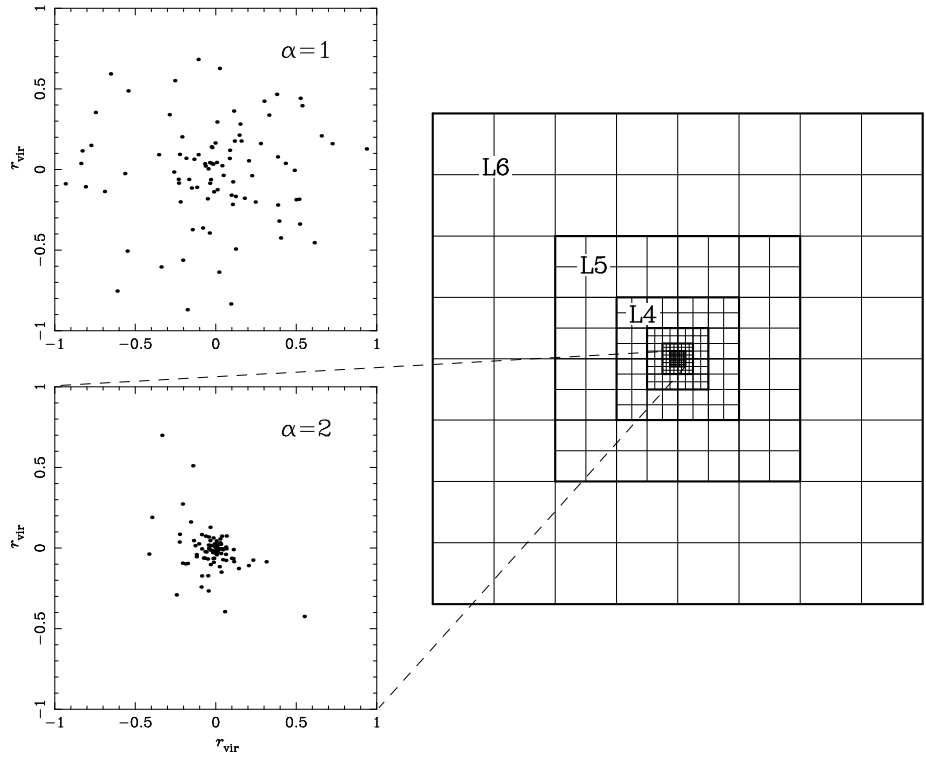}}
\caption{\footnotesize
{\it Left}: Projected distributions of OB associations. The upper and 
lower panels correspond to Case 1 and Case 2, respectively. 
{\it Right}: The structure of the nested grids is shown for a two--dimensional
sectional plane. Six levels of fixed Cartesin grids were used, with the same 
number ($128\times 128\times 128$) of cells for every level.
The box size of the finest grid (L1) was set equal to $2 r_{\rm vir}$, so L1
covers the entire subgalactic halo. 
}
\label{fig4}
\end{figure*}

One of the main purposes of our study is to determine how far 
from the production sites can the product of stellar nucleosynthesis 
propagate into the IGM. Thus, we have to deal with very different length 
scales in our simulation. If we assume, e.g., a minimum resolution 
length comparable to the scale of an individual SN remnant, $\sim 30$ pc, and 
set the size of the simulation box to forty times the virial radius of the 
halo, i.e. 60 kpc for $h=0.5$, then about 2,000 zones are needed
per dimension. Even with massive supercomputers, it is difficult to carry
out such a simulation in three dimensions.
We have therefore adopted a 3--D `nested grid method'.
Our scheme is similar to Tomisaka's (1996) two--dimensional version: 
the general algorithm is based on the works of Berger \& Oliger (1984) and 
Berger \& Colella (1989). Six levels of fixed Cartesian grids were used, with 
every fine grid being completely covered by a coarser one. We named the grids 
L1 (the finest), L2,...., L6 (the coarsest), as shown in the right panel of 
Figure \ref{fig4}. 
The box size of grid L1 was set equal to $2 r_{\rm vir}$, and the mesh size of 
the L$n$ grid to twice that of the L$n-1$ level. All grids were centered 
within each other, with the finest covering the whole galaxy halo. 
Since the cell number is the same ($128\times 128\times 128$) for 
every level L$n$ ($n=1, 2, ..., 6$), the minimum resolved scale is 
about 22 pc and the size of the coarsest grid is 96 kpc. Thus, the scheme has
a wide dynamic range in the space dimension.

The grids are connected by the transfer of conserved variables: the mass 
density $\rho$, the components of the momentum density $\rho\v$, and the total 
energy density $\rho e$. Since each coarse cell is resolved by exactly $2^3$ 
fine cells, the part of the coarse grid covered by the finer grid is overwritten 
with the arithmetic averages of these variables on 8 fine cells. The boundary 
conditions for a fine grid are determined instead by monotonic interpolation of 
physical variables on the coarser grid. For the coarsest level L6 we adopt 
outflowing boundary conditions by imposing for each variable a zero gradient.
Since the Courant--Friedrichs--Lewy condition requires the time--step of 
a coarser grid to be twice as long as the time--step of the next finer level,
the finer grids must be calculated much more often.
The finest grid (L1) is calculated first using an appropriate time--step 
$\Delta t$.  After two time--steps of finest grid, the grid one level coaser 
is advanced by the time--step of $2\Delta t$. The scheme is repeated 
recursively until all grids are advanced. Every time the sequence changes 
from a fine to a coarse grid, the data on the fine grid are averaged onto 
the coarse one.

The center of the fixed NFW dark matter potential is located at the center of 
the simulation boxes, and the halo initial gas distribution follows equation 
(\ref{rhog}). The halo gas is assumed to be non self--gravitating, of 
primordial composition, optically thin, and in collisional ionization 
equilibrium. We assume that the metal ejecta either do not mix with the ambient
gas or do so at late times (cf. Paper I). Even if some mixing occurs inside the
hot cavity gas thereby increasing its metallicity, because of the very low density of this gas 
the cooling rate should remain much lower than in the outer cooling shell. The external IGM is at 
$T=10,000\,$K, as expected from photoheating by the SN progenitors. We use the 
radiative cooling function of 
Sutherland \& Dopita (1993) for primordial gas, and include 
the effect of Compton cooling off microwave background photons, 
\begin{eqnarray} \Lambda_{\rm c}=(5.4\times10^{-36}\,{\rm erg\,cm^{-3}\,s^{-1}})\, \chi\rho\,(1+z)^4\,T.
\end{eqnarray}
Here $\chi$ is the ionized fraction, $z$ (=9) is the 
redshift, and $T$ is the gas temperature. The cooling and heating 
terms are  integrated implicitly.  

Mechanical feedback from SNe is a critical process in galaxy 
formation studies and simulations, as it modifies the composition and 
thermodynamical state of the ambient gas. Different approximate prescriptions 
have been adopted in the past, largely imposed by numerical limitations 
and a poor appreciation of how to implement feedback in hydrodynamical 
simulations (Katz 1992; Navarro \& White 1993; Mihos \& Hernquist 1994; Mori 
et al. 1997; MacLow \& Ferrara 1999; Mori, Yoshii, \& Nomoto 1999). 
Navarro \& White (1993) and Mihos \& Hernquist (1994) assumed that 
a fraction $f_v$ of the available SN explosion energy is deposited in the 
ambient gas as a radial velocity perturbation directed away from the 
event, the remainder being dumped as heat. This method is obviously rather 
sensitive to the assumed value for $f_v$.
Mori et al. (1997) and Mori, Yoshii, \& Nomoto (1999) 
assumed that the gas within the maximum radius of the shock front
in the adiabatic phase of a supernova remnant (Shull \& Silk 1979) 
remains adiabatic until the multiple SNe II phase ends 
at $t_{\rm ms}(8\,\msun$). In this case, the 
effects of radiative cooling might have been be underestimated. 
MacLow \& Ferrara (1999) modeled SN feedback in a central
region of a dwarf galaxy as a constant luminosity wind driven by a
thermal energy source. Nearby starbursts, however, are known
to have multiple star formation sites scattered across the whole galaxy. 

Our algorithm for simulating SN feedback improves upon previous treatments
in several ways. OB associations are distributed as a function of gas 
density ($\propto \rho^\alpha$) using a Monte--Carlo procedure  
(the projected distributions of associations in plotted in Figure \ref{fig4}). 
After a main sequence lifetime, all stars more massive 
than 8 M$_{\odot}$ explode instantaneously injecting an energy of $10^{51}$ 
erg, and their outer layers are blown out leaving a compact remnant of 1.4 
M$_{\odot}$. Therefore SNe inject energy (assumed in pure thermal form) and mass into 
the interstellar medium: these are supplied to a sphere of radius 
\begin{eqnarray}
R_s=24 ~{\rm pc} \left( \frac{\Delta t}{10^4 {\rm ~yr}} \right)^{2/5}
   \left( \frac{n}{0.1 {\rm ~cm}^{-3}} \right)^{-1/5},
\end{eqnarray}
corresponding to the radius of a SN remnant in a uniform ambient medium of density
$n$ and in the adiabatic Sedov--Taylor phase; the expansion velocity $\dot R_s$ 
is also self--consistently calculated from such solution. The gas then starts cooling 
immediately according to the adopted gas cooling function.
The radius $R_s$ has a minimum value of 22 pc due to numerical resolution.
The time--step $\Delta t$ is controlled by the Courant condition. One
additional constraint is that, if there are more than two SN events in 
a OB association, we decrease $\Delta t$ until only one explosion per
association occurs during the time--step $\Delta t$.
The main limitation of our method is the assumption that the gas is 
initially in hydrostatic equilibrium and non self--gravitating.

\section{Numerical Results}

\begin{figure*}
\centerline{
\psfig{figure=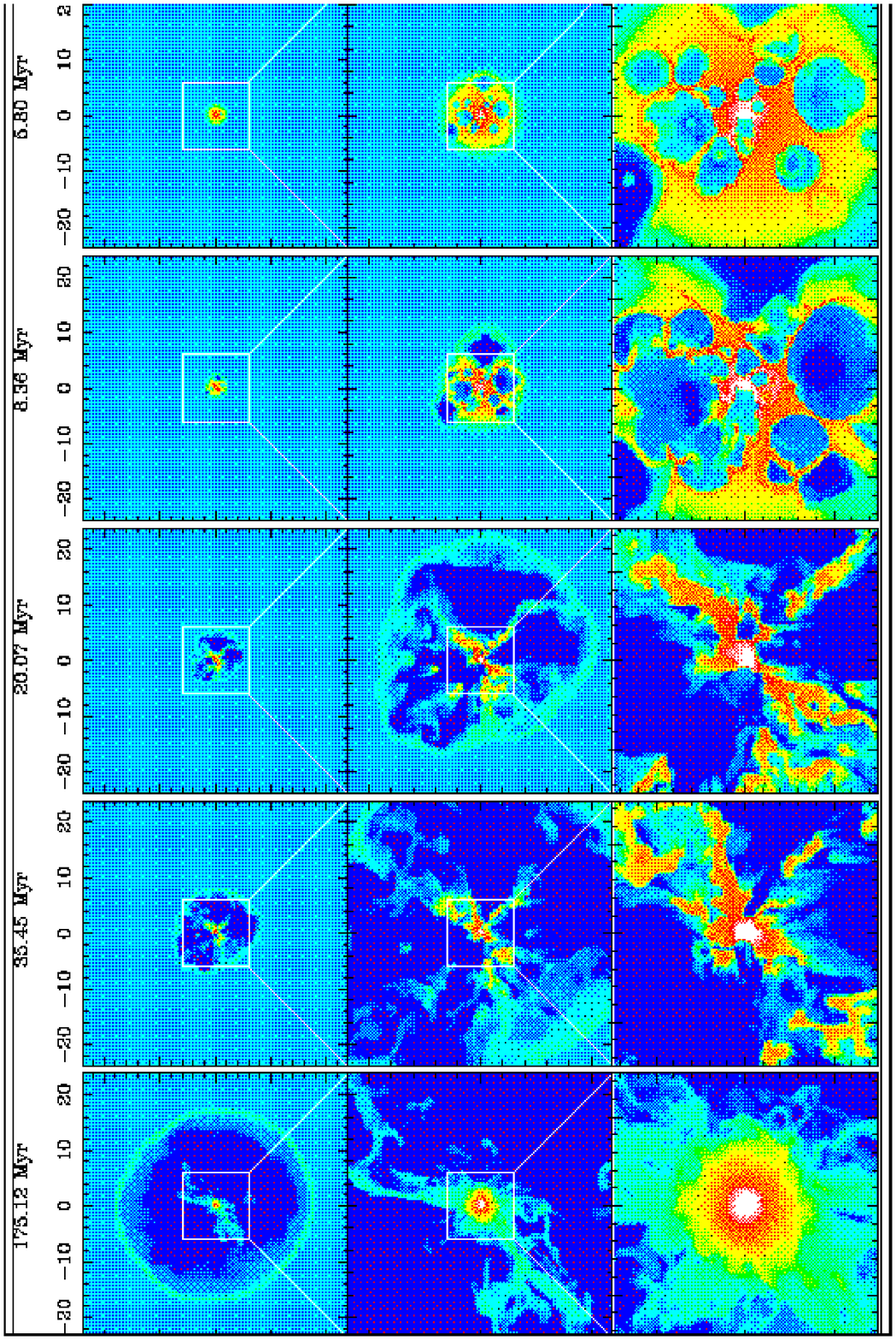,height=20cm}}
\caption{\footnotesize Snapshots of the logarithmic number density of the gas 
at five different elapsed times for our Case 1 simulation run. The three
panels in each row show the spatial density distribution in the $X-Y$ 
plane on the nested grids.
The left, middle, and right panels in each row correspond to the level 
L5, L3, and L1 grid, respectively.
}
\label{fig5}
\end{figure*}

\begin{figure*}
\centerline{
\psfig{figure=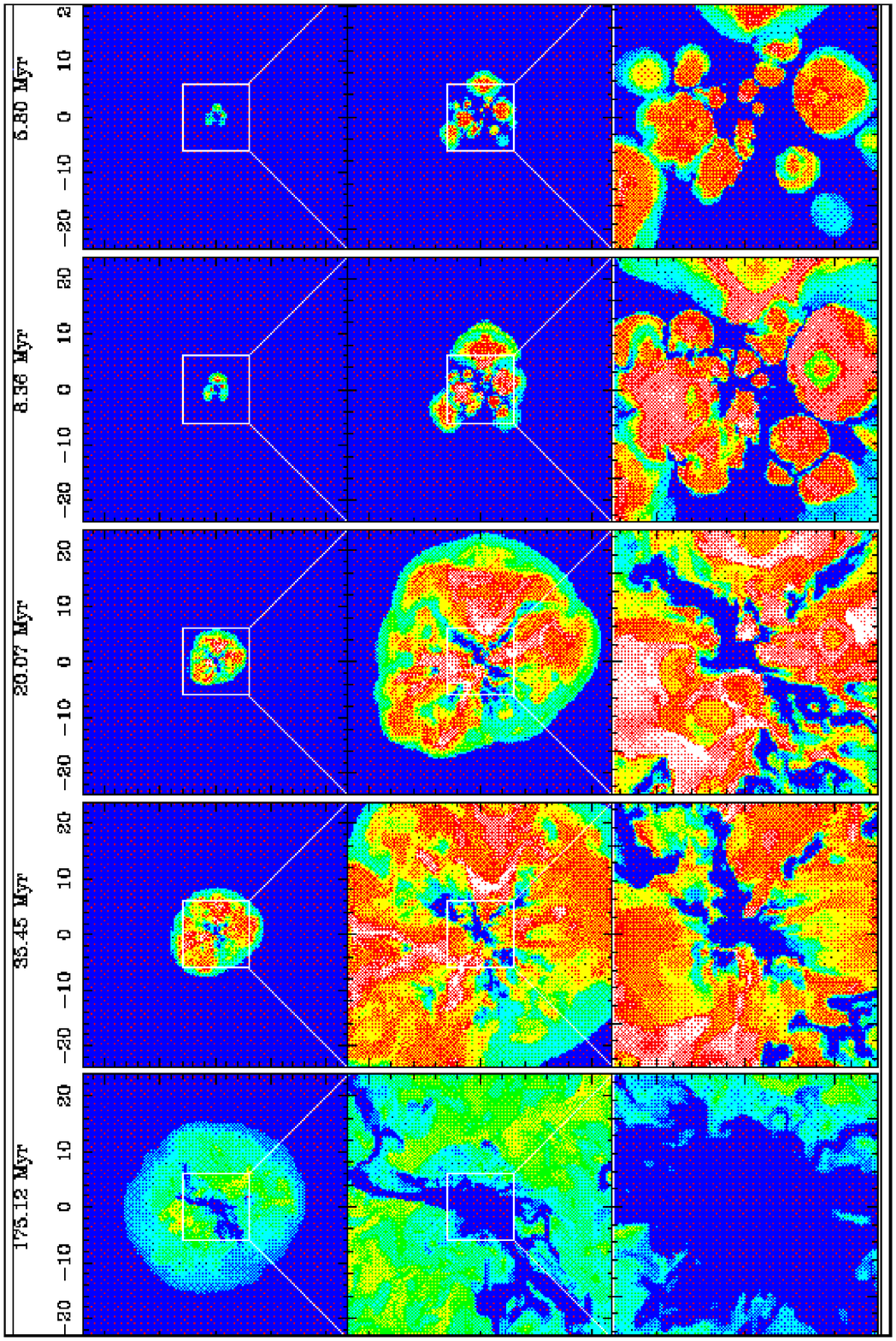,height=20cm}}
\caption{\footnotesize Snapshots of the logarithmic temperature of the gas 
at five different elapsed times for our Case 1 simulation run.
The three panels in each row show the spatial temperature distribution in 
the $X-Y$ plane on the nested grids.
The left, middle and right panels in each row correspond to the level 
L5, L3, and L1 grid, respectively.
}
\label{fig6}
\end{figure*}

\begin{figure*}
\centerline{\psfig{figure=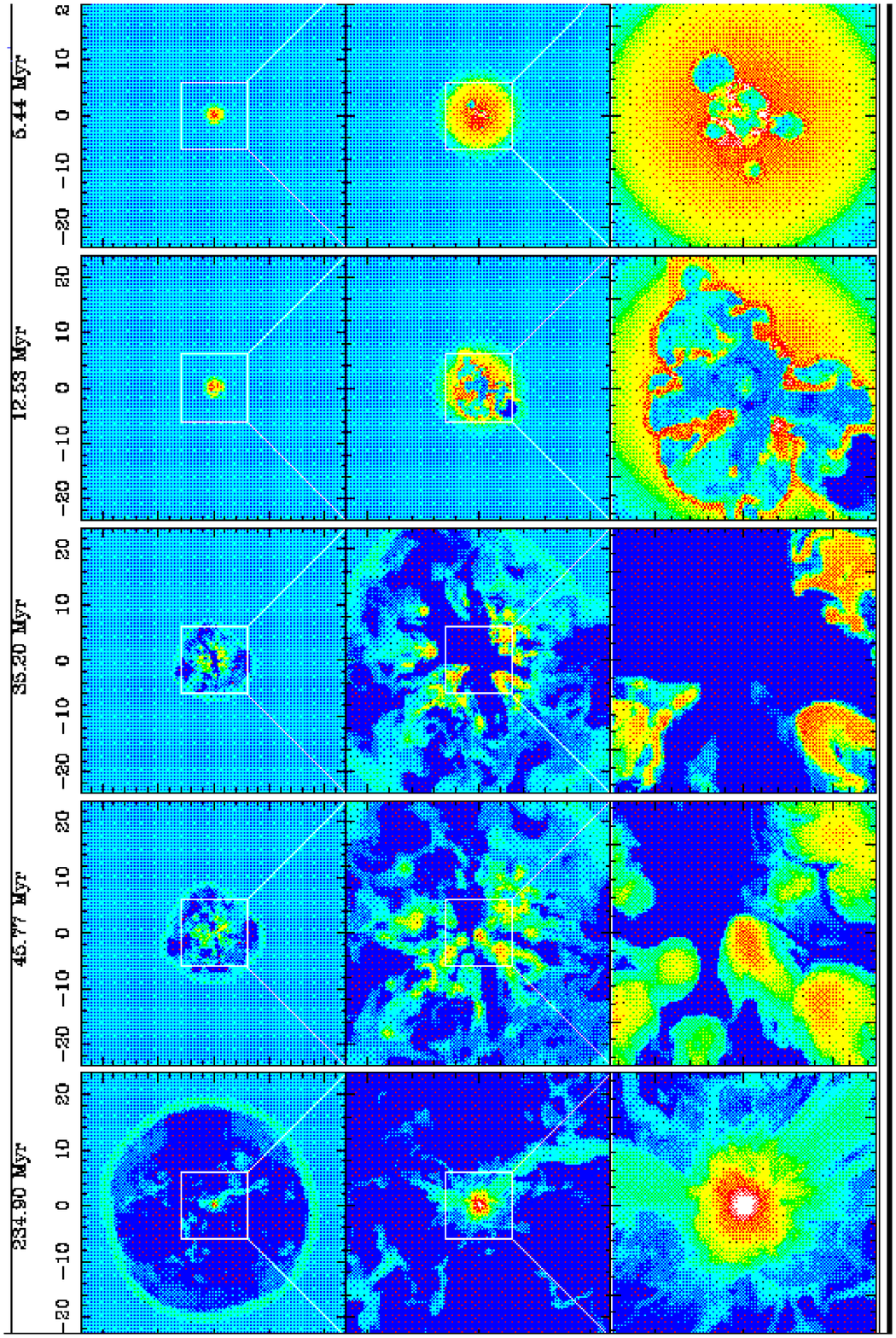,height=20cm}}
\caption{\footnotesize
Same as in Fig. \ref{fig5} but for our Case 2 simulation run.
}
\label{fig7}
\end{figure*}

\begin{figure*}
\centerline{
\psfig{figure=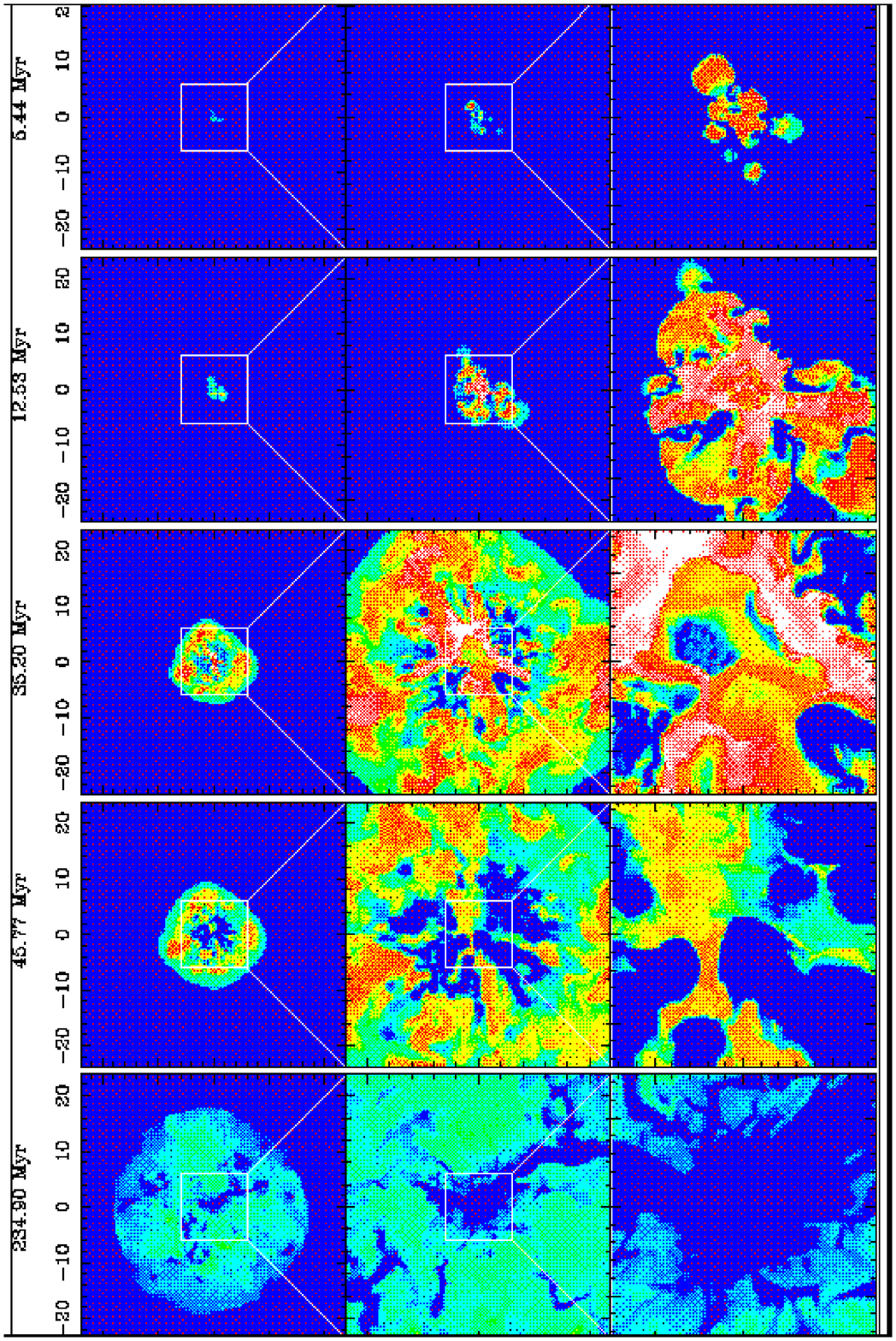,height=20cm}}
\caption{\footnotesize Same as in Fig. \ref{fig6} but for our Case 2 
simulation run.
}
\label{fig8}
\end{figure*}

As mentioned above, we have run two numerical simulations, an extended stellar 
distribution (Case 1) and a more concentrated one (Case 2). In both runs 
11170 SNe explode in our halo (corresponding to $f_\star= 10\%$). Snapshots of the 
gas density and temperature distributions in a slice along the X--Y 
plane of the nested grids are shown in Figures \ref{fig5} and \ref{fig6} 
for Case 1, and Figures \ref{fig7} and \ref{fig8} for Case 2.
The three columns in each figure depict the time evolution from about 5 Myr to 
up to 200 Myr. Along a given row, the leftmost panel refers to grid L5 (linear 
size 48 kpc), the central one to grid L3 (linear size 12 kpc), and the rightmost 
panel refers to the grid L1 (linear size 3 kpc). The density range is  
$-5 \le \log~(n/{\rm cm}^{-3}) \le 1$, and the temperature range is $4 \le 
\log~(T/{\rm K}) \le 8.5$.

After a few Myr from the beginning of the simulation the most massive stars 
explode as SNe and produce expanding hot bubbles surrounded
by a cooling dense ($n \approx 1$~cm$^{-3}$) shell.
At these early times the typical bubble size tends to 
be larger in Case 1 than in Case 2 because in the former scenario the
explosions sites are more uniformly distributed and more likely to occur in 
the outer, lower--density regions, which favors their rapid expansion. 
In spite of the smaller typical bubble size, the degree of overlapping
appears to be more pronounced in Case 2, indicating that the crowding effect is
dominant.

The different initial topology of the multiphase ISM leaves an inprint also
in the later evolutionary stages. At around 10 Myr the ISM structure is relatively
ordered in Case 2, where individual bubbles have merged in a coherent (although
far from spherical) expanding superbubble from which cold ($T\approx 10^4$ K) 
filaments protrude inside the cavity. These filaments are 
the leftovers of previous individual shell--shell interactions further processed
by hydrodynamic instabilities. In Case 1, the halo topology is more 
perturbed, with bubbles expanding in the outer regions 
having already undergone blowout and venting their hot gas into the IGM (see,
e.g., the structure at the top of the panel in the second row, third column of 
Fig. \ref{fig5}), and others whose interaction is giving rise to an intricated, 
multiphase structure in the inner halo,  
where $10^8$~K gas coexists with a cooler $10^4$~K phase from which it is
separated by cooling interfaces. Note that for Case 1 SN explosions in the 
outer halo drive inward--propagating shocks that act to collect and
pile up gas towards the center. This effect is much less pronounced in Case 2 
where the net mass flow is an expanding one. As we will see below, 
the impact of the mechanical energy deposition on the host pregalactic halo
is rather different in the two simulations.

As the evolution continues, a coherent and increasingly spherical shell
expanding into the IGM is eventually formed in both runs. The shell contains 
a large fraction of the halo gas that has been swept--up during the evolution; 
at $t=20\,$ Myr its size is roughly 6 kpc for Case 1 and slightly larger for 
Case 2, as a consequence of the most efficient use of mechanical energy in the 
latter simulation. In Case 1 one can clearly see a central, dense core
resulting from the `implosion' wave  mentioned above. While such a feature is 
almost absent in Case 2 at $t=35\,$ Myr, a dense core will form at later stages 
as a result of the accretion of cold clumps that are balistically raining towards
the center.

\begin{figure*}
\centerline{
\psfig{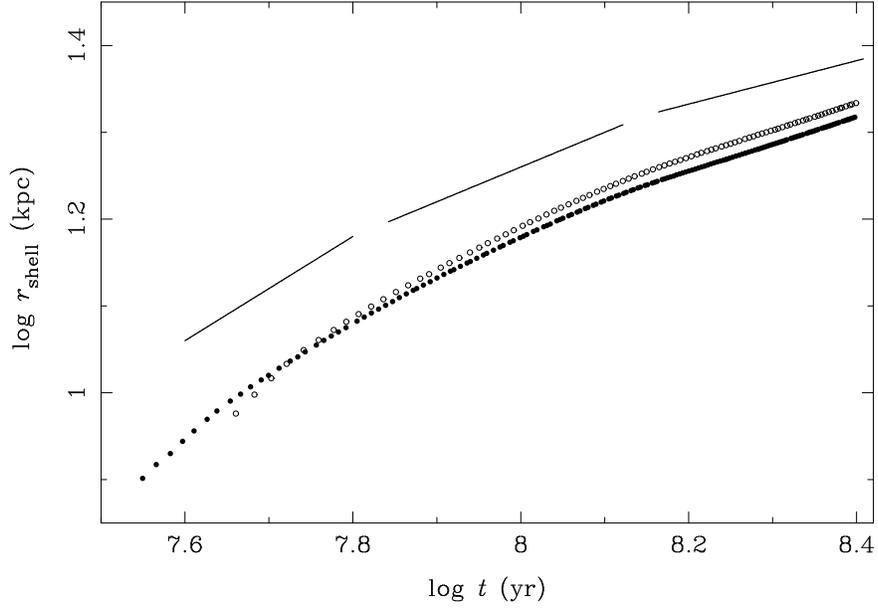}}
\caption{\footnotesize
The evolution of the shell radius as a function of time.
The {\it filled (open) circles} correspond to the Case 1 (Case 2) simulation run.
The {\it solid lines} show the expected power--law behavior in the 
energy--driven phase ($r_{\rm shell} \propto t^{3/5}$), the adiabatic Sedov--Taylor 
solution ($r_{\rm shell}\propto t^{2/5}$), and the momentum--conserving `snowplough' 
phase ($r_{\rm shell} \propto t^{1/4}$). 
}
\label{fig9}
\end{figure*}

The final two bottom rows of the simulation figures show the final stages of 
the evolution that are qualitatively similar to the one just described. 
The shell is now nearly spherically symmetric: its interior is filled with warm 
($T\lta 10^6$~K) 
gas at a very low density $n \simlt 10^{-4}$~cm$^{-3}$, \ie slightly below the 
mean value for the IGM. Figure \ref{fig9} shows the locus of the 
spherically--averaged shell radius $r_{\rm shell}$ as a function of time. 
The shell initially follows an energy--driven phase where 
$r_{\rm shell} \propto t^{3/5}$ (Weaver \etal 1977). At later time the evolution 
relaxes to the adiabatic Sedov--Taylor solution with $r_{\rm shell}\propto t^{2/5}$.
Afterwards the shell slows down to a momentum--conserving `snowplough' phase,
$r_{\rm shell} \propto t^{1/4}$. 
Figure \ref{fig9} shows that, at the end of the simulations, the shell is 
still sweeping  out IGM material; its
radius and  velocity are 21 kpc and 26 km sec$^{-1}$ 
at $t=250\,$Myr. At Mach number ${\cal M}=1$, as the shock will decay into sound 
waves, the maximum `stalling radius' will be reached. 
Using momentum conservation we estimate the final radius of the shell 
to be close to 26 kpc. This is about a factor 2.3 larger than the value
we derived analytically in Paper I for the same case.

\begin{figure*}
\centerline{
\psfig{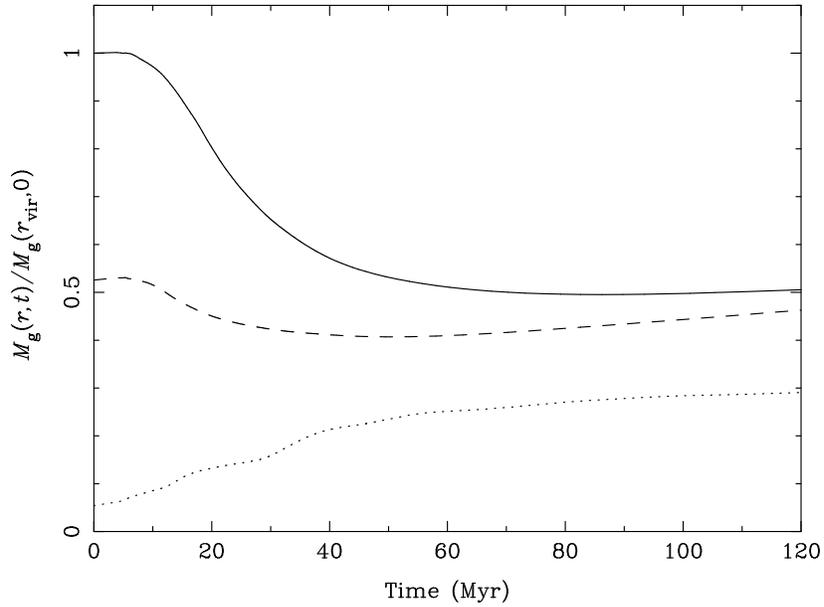}}
\caption{\footnotesize
The evolution of the gas mass inside the gravitational potential 
well of the CDM halo for our simulation run of Case 1 ($\alpha=1$) 
as a function of time.
Curves correspond to the gas mass inside the virial radius 
$r_{\rm vir}$ ({\it solid line}), $0.5 r_{\rm vir}$ ({\it dashed line}), 
and $0.1 r_{\rm vir}$ ({\it dotted line}).
}
\label{fig10}
\end{figure*}

In both simulations runs the final configuration includes
a central core resulting from the two different mechanisms already outlined;
its central density and radial profile is similar to the initial one.
However, the relative ratio between the gas mass contained in the shell and the
one in the central core is different. A visual inspection of 
Figures \ref{fig5} and \ref{fig7} already shows a thicker shell and a 
less massive core in Case 2 relative to Case 1. This is confirmed
quantitatively by the plots in Figure \ref{fig10} (Case 1) 
and Figure \ref{fig12} (Case 2) where we show the fraction 
of the initial halo baryonic mass contained inside $(1, 0.5, 0.1)
r_\vir$ as function of time. The differences are striking: in Case 1, 
the amount of gas at the center is constantly increasing, finally
collecting inside 0.1 $r_\vir$ about 30\% of the total initial mass. 
On the contrary, in Case 2 the central regions remain practically
devoided of gas until 60 Myr, when the accretion process starts. The final 
result is a small core containing a mass fraction of only 5\%.
In the former case 50\% of the halo gas mass is ejected together
with the shell, whereas in Case 2 this fraction is $\sim 85\%$, i.e. the 
blow--away is nearly complete. 

\begin{figure*}
\centerline{
\psfig{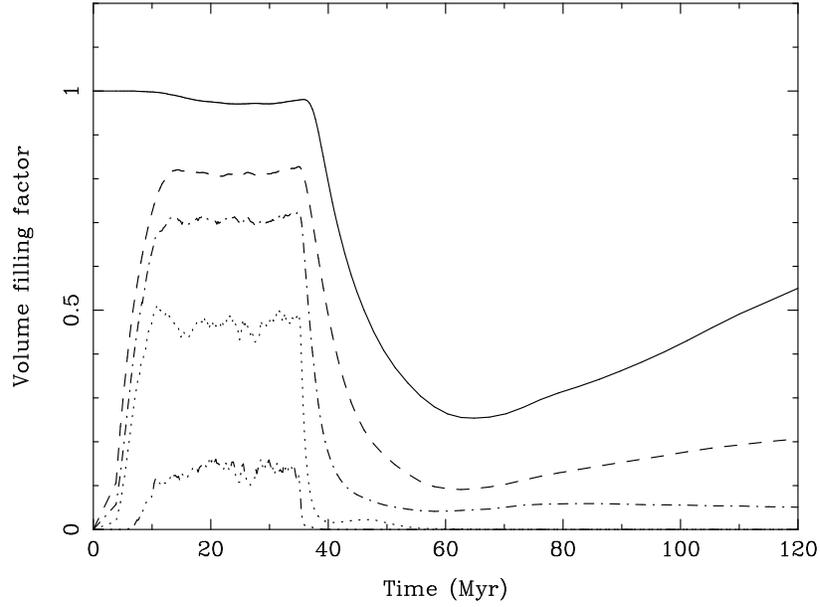}}
\caption{\footnotesize
Cumulative evolution of the volume filling factor inside the virial radius of
the gravitational potential well of the CDM halo for our simulation 
run of Case 1 ($\alpha=1$), as a function of time.
Each curve corresponds to a different gas temperature 
({\it solid line}: $T=10^4$ K,
{\it dashed line}: $T=10^5$ K, {\it dash--dotted line}: $T=10^6$ K, {\it dotted 
line}: $T=10^7$ K, and {\it dash--dot--dotted} line: $T=10^8$ K).
}
\label{fig11}
\end{figure*}

\begin{figure*}
\centerline{
\psfig{figure=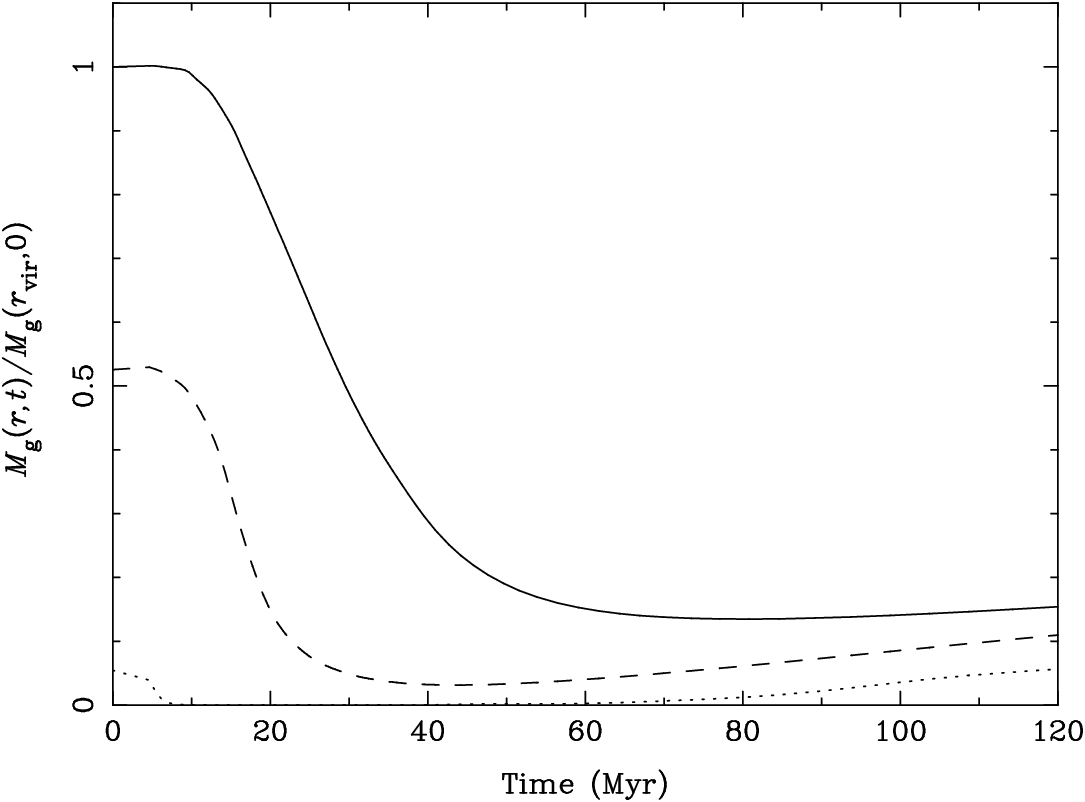,height=8cm,angle=-90}}
\caption{\footnotesize
Same as in Fig. \ref{fig10} but for our Case 2 simulation run.
}
\label{fig12}
\end{figure*}

\begin{figure*}
\centerline{
\psfig{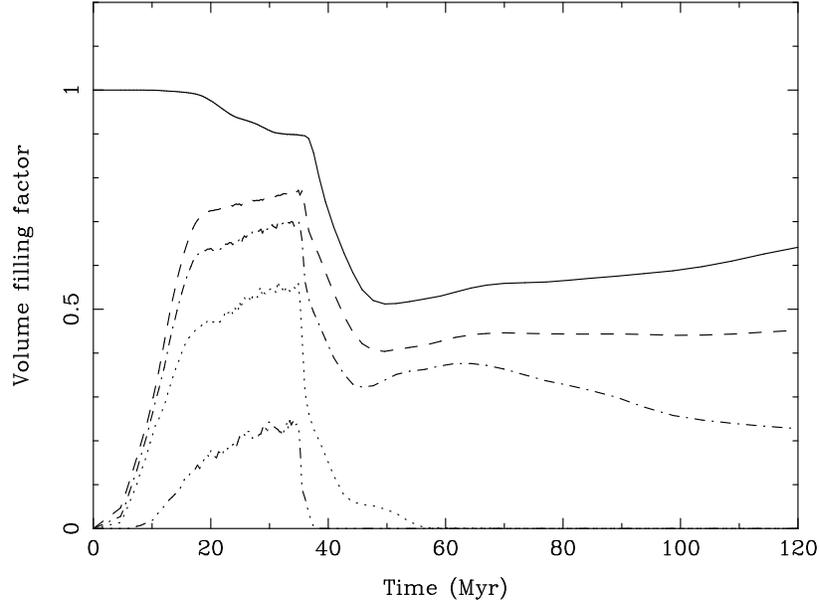}}
\caption{\footnotesize
Same as in Fig. \ref{fig11} but for our Case 2 simulation run.
}
\label{fig13}
\end{figure*}

The gas thermal history is illustrated in Figures \ref{fig11} (Case 1) and 
\ref{fig13} (Case 2), where the evolution of the filling factor of the 
components with temperature larger than a given threshold
[$\log (T/{\rm K}) = 4, 5, 6, 7, 8$] is shown inside $r_\vir$. In both 
simulations we
observe an initial conversion of cold gas into hot gas followed by
the opposite, much slower,  process. The main difference between the two 
runs consists in the larger extension of the warm $T=10^{5-6}$~K component 
filling 20-50\% of the volume in Case 2 at 120 Myr.
Also the compression suffered in the inner regions results for Case 1 in a 
very compact cold gas core, confined to 25\% of the volume by 60 Myr, which 
re-expands afterwards. Case 2 produces a more extended hot gas
distribution as seen from the final values of the curves at 120 Myr.

\begin{figure*}
\centerline{
\psfig{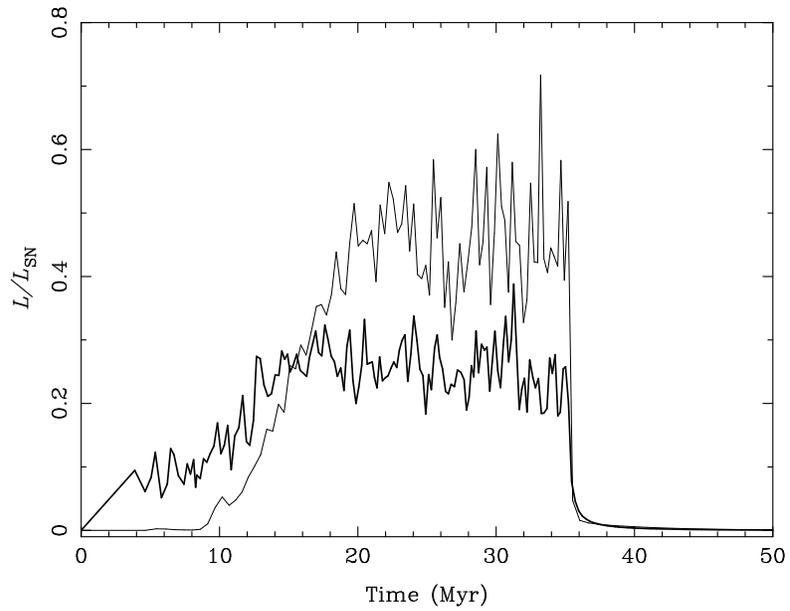}}
\caption{\footnotesize
Fraction of the SN mechanical luminosity carried by the
outflowing gas as kinetic energy through the virial radius, 
as a function of time. The thick line corresponds to the simulation run of Case 1,
and thin line corresponds to that of Case 2.
}
\label{fig14}
\end{figure*}

Figure \ref{fig14} shows the kinetic energy flux, ${\cal E}_k$,  
carried by the  outflowing gas through the 
virial radius.  The kinetic energy flux is calculated as
\begin{eqnarray}
{\cal E}_k=4 \pi r_{\rm vir}^2 \bar{F},
\end{eqnarray}
with
\begin{eqnarray}
\bar{F}= \sum_i dV_i \frac{1}{2} \rho_i {v_r}_i^3 / \sum_i dV_i,
\end{eqnarray}
where $dV_i$ is a volume of the grid cell located at $r_{\rm vir}$ from the 
center, and $\rho_i$ and ${v_r}_i$ are the gas density and the radial 
velocity at that location, respectively.
The summation is taken only for the case of a positive radial velocity
at the virial radius. The mean SN mechanical luminosity, $L_{\rm SN}$, is
defined as
\begin{eqnarray}
L_{\rm SN}&=&\frac{E_{\rm SN}}{t_{\rm ms}(8M_\odot)-t_{\rm ms}(120M_\odot)},\\
  &=&1.1 \times 10^{40}    {\rm ~~~erg ~s^{-1}}.
\end{eqnarray}
From Figure \ref{fig14}, we see that kinetic energy is ejected from
the galaxy at roughly a constant fraction 25\% of the  
SN mechanical luminosity from 13 Myr to 35 Myr for the Case 1 run; 
this value increases to about 40\% (but with a larger time spread) from 13 Myr 
to 35 Myr for the Case 2 run. Averaged over the entire evolution, we
find that 23\% (30\%) of the total SNe energy is converted to the kinetic 
energy of the outflowing gas for the run of Case 1 (Case 2). 

\section{Summary and discussion}

In this Paper we have used three--dimensional hydrodynamic simulations to 
investigate the dynamics of SN--driven bubbles as they propagate through and escape 
the grasp of subgalactic halos with masses $M=10^8\,h^{-1}\,\msun$ at redshift $z=9$. 
Depending on the stellar distribution, we found that less than 30\% of the available 
SN energy is converted into kinetic energy of the blown away material, and that 
mechanical feedback is less efficient than expected from simple energetic arguments, 
as off--nuclear SN explosions drive inward--propagating shocks that tend to 
collect and pile up cold gas in the central regions of the host halo. 
For the more extended star formation Case 1, the implosion wave collects more
material in the center, eventually resulting in a more massive cold
core in which star formation can be restarted. 
The amount of kinetic energy ejected into the IGM is essentially double for Case 2, 
again because less energy flows towards the center in this run. 
Blow--away is more efficient if stars form preferentially at the center of the 
halo.

SN--driven pregalactic outflows may be an efficient mechanism
for distributing the product of stellar nucleosynthesis over large cosmological
volumes prior to the reionization epoch. 
In Paper I we used several approximations to show that, for
a star formation efficiency of $f_\star=10\%$, the radius of the SN--driven
bubble around our prototypical early halo would grow up with time up to
a final stalling value of about 11 kpc, when pressure equilibrium with the 
IGM was achieved. In this simulations we find larger final radii, up
to about 26 kpc for both runs.
What is the expected spatial extent of the ensemble of wind--driven ejecta from 
a population of pregalactic systems? Consider an Einstein--de Sitter cosmology 
with $\Omega_M=1$, $h=0.5$, $\sigma_{8}=0.63$, $n=1$, and $\Omega_bh^2=0.019$,
where the amplitude of the power spectrum has been fixed in order to reproduce 
the observed abundance of rich galaxy clusters in the local universe (we have
used the transfer function by Efstathiou, Bond, \& White 1992 with shape 
parameter $\Gamma=0.5$). According to the Press--Schechter formalism, the
comoving abundance of collapsed dark halos with masses $M\approx 10^8\,h^{-1}\,
\msun$ at $z=9$ is then close to $80\,h^3\,$Mpc$^{-3}$, corresponding to a mean
proper distance between neighboring halos of $15\,h^{-1}\,$kpc, a total mass 
density parameter of order 3 percent, and a stellar density parameter of 
$0.002\,f_\star$. A $\Lambda$CDM cosmology with the same normalization would give 
a halo number density a factor 3.7 lower, resulting in a larger mean separation
of 23 $h^{-1}$ kpc.
Therefore, while with $f_\star=10\%$ only a small fraction, about
4\% percent, of the total stellar mass inferred at the present epoch
(Fukugita, Hogan, \& Peebles 1998) would actually form at these early epochs,
the impact of such an era of pregalactic outflows could be quite significant, as
the product of stellar nucleosynthesis would be distributed over distance that 
are comparable to the mean proper distance between neighboring low--mass systems,
i.e. volume filling factors of order unity.\footnote{We note that, on this 
assumption, $f_\star$, and hence the early luminosity of these systems, would exceed 
the value predicted by the usual low-mass extrapolation of  CDM galaxy formation 
models (cf. White \& Frenk 1991).
In these scenarios $f_\star$ is postulated to decline steeply in
shallow potential wells, thereby reducing the population of low--luminosity
galaxies and avoiding the so--called `cooling catastrophe'.}\,
The collective explosive output
of about ten thousands SNe per $M\gta 10^8\,h^{-1}\,\msun$ halo at these early
epochs could then pollute the entire intergalactic space to a mean metallicity
$\langle Z\rangle=\Omega_Z/\Omega_b\gta 0.003$ (comparable to the levels
observed in the \lya forest at $z\approx 3$) without much perturbing the IGM
hydrodynamically, i.e. producing large variations of the baryons relative
to the dark matter.

We want to comment here on the possible effect of gas self--gravity, 
neglected in our simulations. The local gas--dynamical time of 
cold filaments and blobs in the halo ISM is of order $3\times10^7$ yr for a gas 
number density of 1 cm$^{-3}$, and $9\times10^6$ yr 
for a density ten times greater. This timescale is comparable 
to the characteristic timescale of gas removal from the potential well of the
dark matter halo in our simulations. If the density of the cold component is 
increased due to self--gravity, radiative cooling will be enhanced. 
This will decrease the efficiency of the conversion of the available SN energy into 
the kinetic energy of the outflowing material, weakening the blow--away.
Two compensating effects -- that act to prevent gas cooling -- should also be 
considered then, photoionization by UV radiation and thermal conduction between 
the cold and hot interstellar medium. In particular, the conduction timescale is
$$
\tau_{\rm cond}=\frac{3nkT}{2\nabla \cdot (\kappa T^{5/2} \nabla T)}
$$
\beq
~~~\approx 4.3~{\rm Myr} \left( \frac{n}{1 {\rm ~cm}^{-3}} \right)
                         \left( \frac{l}{100 {\rm ~pc}} \right)^{2} 
                         \left( \frac{T}{10^7 {\rm ~K}} \right)^{-5/2},
\eeq
where $\kappa$ is the conduction coefficient, $n$ is the gas number density, and
$l$ is the characteristic scale length of the cold filaments. Taking $T=10^7\,$K, 
$l=100$ pc, and $n=1$ cm$^{-3}$ from the results of our simulations,
we find $\tau_{\rm cond}=4.3\times10^6$ yr.
This timescale is shorter than the free--fall time and comparable to the
cooling timescale. The thermal energy of the cold phase will then be affected by 
thermal conduction from the hot medium.

Finally, while in this paper we simulated the effect of an instantaneous burst of
star formation in a protogalactic spherical halo embedded in a uniform IGM, in a 
series of forthcoming studies we will investigate non--instantaneous stellar 
feedback in continuous star forming halos of different masses and
morphologies, within a realistic cosmological density field. In these cases 
the blow--away history of the gas may be quite different.

\acknowledgments
\noindent We thank F. Palla, M. Rees, and M. Umemura for useful discussions. 
This work was started as one of us (A.F.) was a visiting professor at the 
Center for Computational Physics, Tsukuba University. Support for this project
was provided by NASA through ATP grant NAG5--4236 and grant AR--06337.10-94A 
from the Space Telescope Science Institute, and by a B. Rossi visiting 
fellowship at the Observatory of Arcetri (P.M.). A.F. and P.M. also 
acknowledge the support of the EC RTN network ``The Physics of the 
Intergalactic 
Medium''. The numerical computations were partly carried out on a Hitachi 
SR8000/8 and a PILOT--3/64
at the Center for Computational Physics, University of Tsukuba, and partly on a 
Fujitsu VPP--300 at the National Astronomical Observatory in Japan.

\appendix

A number of different tests were performed to demonstrate the ability of our code
to reproduce known analytical results. Here we present results for: (i) a 
standard one--dimensional shock tube, and (ii) the adiabatic expansion of a point 
explosion in three dimensions. The hydrodynamic equations (\ref{compact}) were 
solved with the source term {\bf S} set equal to zero, on a grid of 128 zones in 
each dimension. 

\begin{figure*}
\centerline{
\psfig{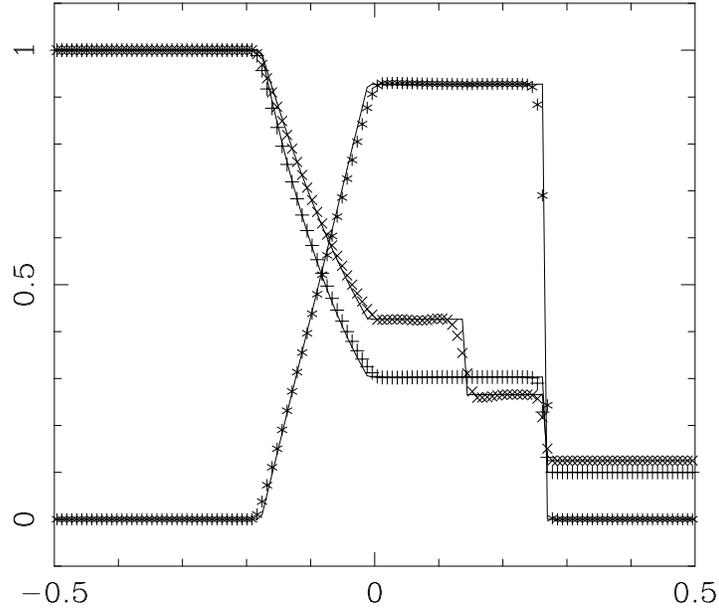}}
\caption{\footnotesize Normalized density ({\it crosses}), pressure
({\it pluses}), and velocity ({\it asterisks}) profiles in a one--dimensional 
shock tube. The corresponding analytical solutions are shown by the {\it solid 
lines}.
}
\label{figA1}
\end{figure*}

In the shock tube two regions of different gas densities are instantaneously brought
into contact (see Sod 1978). A shock (rarefaction) wave propagates then into the 
low (high) density gas. We choose the initial conditions to be:
$$
\begin{array}{lllll}
\rho=1,& P=1,& v=0,& {\rm ~~~for}& x \leq 0; \\
\rho=0.125,& P=0.1,& v=0,&   {\rm ~~~for}& x > 0. \\
\end{array} \eqno(\rm A1)
$$
The ratio of specific heats is $\gamma=1.4$ and the shock Mach number is 1.66.
In this test we integrate the hydrodynamic equations on a fixed grid, 
\ie without the nested 
grid method. The numerical results for the density, pressure, and velocity are 
shown in Fig. \ref{figA1} and compared with the analytical solutions. The code 
can resolve a shock front within 2 cells, while contact discontinuities are 
spread over 3 cells. Note how the AUSMDV scheme suppresses oscillations in the 
postshock velocity profile (cf. Cen 1992).

\begin{figure*}
\centerline{
\psfig{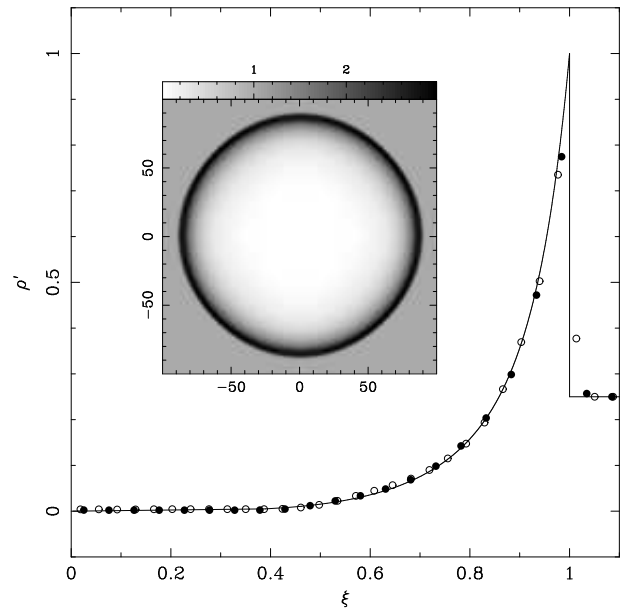}}
\caption{\footnotesize Point explosion in a homogeneous medium. 
The normalized density profile is shown as a function of the 
(self--similar) radius at elapsed times of 2638 yr ({\it filled circles}) 
and 10382 yr ({\it open circles}). The self--similar Sedov--Taylor solution for 
the gas density is indicated by the {\it solid line}. The 
inlet shows a two--dimensional cut through the explosion site.
}
\label{figA2}
\end{figure*}

The second test run follows the evolution of a spherical blast wave caused by a point 
explosion in a homogeneous gas with $\gamma=5/3$, neglecting radiative cooling. 
Two grid levels 
(L1 and L2) are used in this case. The gas density is set to $\rho_i=10^{-24}$ gr 
cm$^{-3}$ and the gas temperature to $T_i=100$ K. An explosion energy of
$E_i=10^{51}$ ergs is added to the eight zones at the center of the L1 grid. The 
propagation of the shock front is described by the self--similar Sedov--Taylor 
solution $r_{\rm shell}=\xi_0 (E_i/\rho_i)t^{2/5}$, where the coefficient $\xi_0$ 
is 1.15 for $\gamma=5/3$. Fig. \ref{figA2} shows the normalized density 
profile as a  
function of radius at elapsed times 2638 yr and 10382 yr. The shock wave propagates 
outward from the center of the L1 grid, and smoothly passes through the boundary 
between grid levels L1 and L2. The numerical results nicely reproduce the analytic 
shock position and profile.

\end{document}